\documentclass[aps,prd,twocolumn,showpacs,superscriptaddress,groupedaddress,nofootinbib]{revtex4-1}
\usepackage{dcolumn}   
\usepackage{bm} 
\usepackage{amssymb}
\usepackage{graphicx}
\usepackage{subfigure}
\usepackage{epsfig}
\usepackage[tbtags]{amsmath}
\usepackage{exscale}

\begin{document}

\setcounter{footnote}{0}

\newcommand{\lp}{\ell_{\mathrm P}}

\newcommand{\md}{{\mathrm{d}}}
\newcommand{\tr}{\mathop{\mathrm{tr}}}
\newcommand{\sgn}{\mathop{\mathrm{sgn}}}

\newcommand*{\R}{{\mathbb R}}
\newcommand*{\N}{{\mathbb N}}
\newcommand*{\Z}{{\mathbb Z}}
\newcommand*{\Q}{{\mathbb Q}}
\newcommand*{\C}{{\mathbb C}}

\newcommand{\gpp}{g_{\phi\phi}}
\newcommand{\grr}{g_{rr}}
\newcommand{\bgpp}{\bar{g}_{\phi\phi}}
\newcommand{\bgrr}{\bar{g}_{rr}}
\newcommand{\ppp}{\pi^{\phi\phi}}
\newcommand{\prr}{\pi^{rr}}
\newcommand{\bppp}{\bar{\pi}^{\phi\phi}}
\newcommand{\bprr}{\bar{\pi}^{rr}}
\newcommand{\tppp}{\tilde{\pi}^{\phi\phi}}
\newcommand{\tprr}{\tilde{\pi}^{rr}}
\newcommand{\kp}{\ensuremath{K_\varphi}}
\newcommand{\kr}{\ensuremath{K_r}}
\newcommand{\nr}{N^{r}}
\newcommand{\er}{\ensuremath{E^r}}
\newcommand{\ep}{\ensuremath{E^\varphi}}
\newcommand{\gppd}{\dot{g}_{\phi\phi}}
\newcommand{\grrd}{\dot{g}_{rr}}
\newcommand{\epd}{\dot{E}^{\varphi}}
\newcommand{\erd}{\dot{E}^{r}}
\newcommand{\gp}{\ensuremath{\Gamma_\varphi}}
\newcommand{\mr}{\ensuremath{\sqrt{\frac{2M}{r}}}}
\newcommand{\mrl}{\ensuremath{\sqrt{2Mr}}}
\newcommand{\be}{\begin{equation}}
\newcommand{\ee}{\end{equation}}
\newcommand{\bea}{\begin{eqnarray}}
\newcommand{\eea}{\end{eqnarray}}
\newcommand{\dif}{\mathrm{d}}
\newcommand{\ag}{a_{\gamma}}
\newcommand{\kpb}{\ensuremath{\bar{K}_\varphi}}
\newcommand{\krb}{\ensuremath{\bar{K}_r}}
\newcommand{\erb}{\ensuremath{\bar{E}^r}}
\newcommand{\epb}{\ensuremath{\bar{E}^\varphi}}
\newcommand{\erbd}{\ensuremath{\dot{\bar{E}}^{r}}}
\newcommand{\krbd}{\ensuremath{\dot{\bar{K}}_{r}}}
\newcommand{\kpd}{\ensuremath{\dot{\bar{K}}_{\varphi}}}
\newcommand{\gpb}{\ensuremath{\bar{\Gamma}_\varphi}}
\newcommand{\ab}{\bar{\alpha}}
\newcommand{\aone}{\alpha_{1}}
\newcommand{\at}{\alpha_{2}}
\newcommand{\ath}{\alpha_{3}}
\newcommand{\afo}{\alpha_{4}}
\newcommand{\af}{\alpha_{5}}
\newcommand{\erbp}{\bar{E}^{r'}}
\newcommand{\erp}{E^{r'}}
\newcommand{\erpp}{E^{r''}}
\newcommand{\epp}{E^{\varphi'}}
\newcommand{\nrp}{N^{r'}}
\newcommand{\erbpp}{\ensuremath{\bar{E}^{r''}}}
\newcommand{\bo}{b_{1}}
\newcommand{\bt}{b_{3}}
\newcommand{\erbs}{\bar{E}^{r^{2}}}

\title{\Large Midisuperspace quantization: possibilities for fractional and emergent spacetime dimensions}

\author{Rakesh Tibrewala} 
\email{rtibs@lnmiit.ac.in}
\affiliation{Centre for High Energy Physics, Indian Institute of Science,
Bangalore-5, India}
\altaffiliation{Present address: Department of Physics, The LNM Institute of Information Technology, Jaipur-302031, India}

\begin{abstract}
Recently, motivated by certain loop quantum gravity inspired corrections, it was shown that for spherically symmetric midisuperspace models infinitely many second derivative theories of gravity exist (as revealed by the presence of three arbitrary functions in the corresponding Lagrangian/Hamiltonian) and not just those allowed by spherically symmetric general relativity. This freedom can be interpreted as the freedom to accommodate certain quantum gravity corrections in these models even in the absence of higher curvature terms (at a semi-classical level, at least). For a particular choice of the arbitrary functions it is shown that the new theories map to spherically symmetric general relativity in arbitrary number of (integer) dimensions thus explicitly demonstrating that when working with midisuperspace models, one loses the information about the dimensionality of the full spacetime. In addition, it is shown that these new theories can accommodate scenarios of fractional spacetime dimensions as well as those of emergent spacetime dimensions -- a possibility suggested by various approaches to quantum gravity. 
\end{abstract}

\maketitle

\section{Introduction}

Symmetry reduced models of gravity, the minisuperspace \cite{Misner-Mixmaster, Misner-Magic} and the midisuperspace models \cite{Torre-Midisuperspace, Barbero-Midi-LivingRev} as they are often called, play a key role in the investigation of various approaches to quantum gravity. Because of the imposition of spacetime symmetries - homogeneity of space in the case of minisuperspace models and spherical symmetry for (one class of) midisuperspace models - the resulting symmetry reduced Hamiltonian of general relativity is simpler than that of the full theory. It is then hoped that the resulting simplified theory (or model) will be easier to quantize and will provide useful hints for the quantization of the full theory of general relativity.

However, this simplification of the theory for symmetry reduced models comes with a cost; one loses information about the dimensionality of the full spacetime, specifically the dimensions corresponding to symmetry directions. For instance, irrespective of the number of spacetime dimensions, the phase space of the homogeneous and isotropic  Friedmann-Robertson-Walker (FRW) cosmological model is two dimensional, coordinatized by the scale factor and its conjugate momentum. Similarly, the phase space of a spherically symmetric spacetime is $4\infty$ dimensional (corresponding to two metric variables and their conjugate momenta) irrespective of the number of spherically symmetric directions.

This fact, though obvious when stated as above, is not always fully appreciated and this is so for a good reason. The symmetry reduced or midisuperspace models that one considers are obtained by imposing appropriate spacetime symmetries on Einstein's theory (or its higher derivative generalizations like the Lovelock theory \cite{Lovelock}). The corresponding Lagrangian (Hamiltonian) has a definite structure with the information about the dimensionality of the full spacetime present in the coefficients of the various terms in the Lagrangian (Hamiltonian) of the symmetry reduced theory. The information about the dimensionality of the spacetime can be easily retrieved from the solution of the corresponding equations of motion. For instance, the Schwarzschild solution in $n+2$ dimensions ($n>1$ and $n\in\mathbb{Z}$ and where $n$ corresponds to the number of spherically symmetric directions) is of the form $1-c/r^{n-1}$ and thus the form of the solution contains the information about the dimensionality of the full spacetime.       

However, recently it was found that as far as spherically symmetric midisuperspace models are concerned, there is much more freedom in the structure of the theory and the symmetry reduced version of general relativity is just one among infinitely many possible second derivative theories for these spacetimes \cite{NSDTGSSS}. This infinite freedom is reflected in the presence of three (two) arbitrary functions of the metric coefficient $\gpp$ (labeled $\er$ in the bulk of the paper) in the Hamiltonian of the theory in the presence (absence) of the cosmological constant. For specific functional forms of these functions, one can then obtain new spherically symmetric solutions not present in general relativity. The arguments of the previous paragraph then imply that, in general, it would not be possible to tell the dimensionality of the full spacetime to which these solutions correspond to (see section II and mainly section III below).

All this might seem inconsequential and insignificant as these new theories do not correspond to symmetry reduced classical general relativity. This would be the case if not for the quantum theory. As mentioned earlier, one of the main interests in symmetry reduced models is that they provide a simplified setting in which to understand various aspects of quantum gravity. And as is true for any quantum theory, quantum gravity will lead to quantum corrections of classical Einstein theory (including its symmetry reduced versions). These quantum corrections can appear in various forms like mass (and charge) renormalization in the case of Schwarzschild (Reissner-Nordstr\"om) black holes  as well as (quite generally) these should also show up in the form of field renormalization. 

Now, in general, the form of quantum corrections is difficult to constrain and symmetry arguments are often required to obtain a viable quantum theory. The underlying symmetry of general relativity is diffeomorphism covariance and it is expected that the smooth differential geometry structures of classical general relativity will give way to discrete structures in quantum gravity for which the usual notions of diffeomorphism invariance might not directly apply. In such a scenario an algebraic notion of symmetry would be more useful. Such a notion is provided in the canonical formulation of general relativity where general covariance of the theory is encoded in the closure of the constraint algebra (of the Hamiltonian and diffeomorphism constraints) \cite{Hojman-Kuchar-Teitelboim}. It is expected that even when the classical notions of geometry are not applicable, the commutator of the operators corresponding to the classical Hamiltonian and diffeomorphism constraints would close just as in the classical theory.   

In essence, precisely this criterion was used to obtain the new theories referred to above (although these theories, being classical, were obtained by working with the classical phase space variables and not the corresponding operators on the Hilbert space of the symmetry reduced theory). In other words, the criteria for terming the new theories as theories of gravity is that they are diffeomorphism invariant in the $t-r$ plane as revealed by the fact that these theories have a Hamiltonian constraint and a diffeomorphism constraint and these constraints obey the standard constraint algebra of (symmetry reduced) general relativity. The arbitrary functions appearing in the Hamiltonian of the theory can be thought of as corresponding to the possibility of accommodating some of the quantum corrections to spherically symmetric general relativity (without requiring the incorporation of higher curvature/derivative terms).

In the semi-classical regime, where one expects the notion of the metric to be meaningful, field renormalization would presumably show up in the form of a modified metric function (see \cite{GambiniPullin-CompleteSpaceTime, Modesto-SpaceTimeStructure, Husain-ModifiedGR, ModifiedHorizon, Tibrewala-Einstein-Maxwell} for instance, where effects of certain loop quantum gravity (LQG) inspired corrections were considered for Schwarzschild and Reissner-Nordstr\"om geometries). And it is precisely here that the considerations of the previous paragraphs become important as these immediately imply that if, as in the classical theory, one continues to extract the information about the dimensionality of the full spacetime from the form of the metric even in the semi-classical regime then in the presence of non-trivial quantum corrections one would no longer be able to tell the dimensionality of the original spacetime one started with. 

Not only this, but depending on the exact form of the quantum corrections, it might turn out that in the quantum theory the dimensionality of the full spacetime is different from that of the classical spacetime being quantized. Furthermore, generically it will turn out that the dimensionality of spacetime is not even integral but is fractional. To take a concrete example, if a quantization of the Schwarzschild spacetime (in $n+2$ ``classical" dimensions) leads to a metric coefficient with the leading order form $1+cf(r)/r^{n-1}$ with $f(r)$ corresponding to quantum correction then, in certain cases depending on the form of $f(r)$, the spacetime may well be regarded as having fractional dimensions (see section IV below).

This apparent drawback of the theory can, in fact, be turned into a virtue since such a possibility gels well with the scenario of fractional and/or emergent spacetime dimensions as is suggested by various theories of quantum gravity. In a lot of these theories it is the spectral dimension of the quantum spacetime that is calculated like in the causal dynamical triangulation (CDT) \cite{Loll-CDT} or in LQG \cite{CalcagniDiffusion, CalcagniSpectralLQG, CalcagniDimFlow} (see \cite{ModestoFractal} for earlier work in the context of LQG and \cite{ModestoFractalSpinfoam} in the context of spinfoams) as well as in some other approaches like asymptotic safe gravity \cite{ReuterFractalAsym, ReuterLivRev} and Horava-Lifshtiz gravity \cite{HoravaSpectralHL} including the more recent suggestion for fractal spacetime \cite{Calcagni-Fractal} (also see \cite{CarlipSmallStr, CarlipSpontaneous} for a general overview of various approaches to spectral dimension and certain other considerations). The notion of emergent spacetime is also suggested by considerations of gauge-gravity duality \cite{SeibergEmergent, Polchinski-Dualities} (also see \cite{BenedettiSpectral, AfshordiEmergentStochastic, Gudder-CausalSet} for some other scenarios in which notions of fractal or emergent spacetime show up).

At this point it is worth noting that the notion of spectral dimension mostly probes the microstructure of the spacetime and is not directly related to the dynamics of the underlying theory. However, one expects that in high curvature regime quantum gravity effects will modify the classical dynamics. There then exists the possibility that the modified dynamics might lead to a dynamical evolution of spacetime dimensions. It is the second possibility that is examined in this paper. We will motivate the notion of emergent spacetime from the perspective of effective quantum gravity by constructing an example where the exponent $n$ in $r^{n}$ is not a constant but is a function of the scale (say, $n\equiv n(\ell_{P}/r)$, $\ell_{P}$ being the Planck length), so that only in the classical limit $r\gg\ell_{P}$ do we recover the classically observed dimensionality of spacetime.

Although the specific construction presented here for a symmetry reduced model would not extend exactly to the full theory, it might nevertheless be useful to consider such toy models to develop an understanding and intuition about how the new features of quantum gravity might appear at an effective level. It might be thought that a theory based on conventional notions of spacetime and differential geometry cannot really lead to an emergent notion for spacetime and that one would need a different (non-gravitational) starting point as is the case with matrix models, CDT as well as scenarios involving gauge-gravity duality. While this might be true of the full theory of quantum gravity, we also believe that if there is a notion of spacetime dimensions emerging as a function of some scale in the quantum theory then it should also be possible to capture this feature at some kind of an effective level.

It should be realized that spacetime is not a directly observable entity and that the dimension of spacetime appears only in the dynamics, and since the dynamics may be modified away from the classical limit, the dimension may be modified as well in an effective model. In other words, the present work is a new perspective on dimension suggested by effective models.

Below we elaborate on the ideas suggested above. We work in ``$2+1$" dimensions -- the lowest spacetime dimensions in which spherical symmetry ansatz makes sense (the quotes highlighting the fact, as will be demonstrated below, that for symmetry reduced models the  meaning of dimensionality is non-trivial). In the next section we present the new Hamiltonian(s) (incorporating the three arbitrary functions mentioned earlier). The arbitrariness of these functions is used in section III to construct a new solution for a \emph{supposedly} $2+1$ dimensional spacetime. This solution is used  to demonstrate that the concept of the dimensionality of full spacetime is not so easy to address in midisuperspace models by showing that the solution so obtained actually corresponds to some higher dimensional solution in general relativity. To remove any ambiguity regarding the point being made we further show that with an appropriate choice for the arbitrary functions, the Hamiltonian of the (supposedly $2+1$ dimensional) non-Einsteinian midisuperspace model can be mapped to the Einsteinian (general relativistic) midisuperspace model in $n+2$ dimensions. In section IV we suggest possible implications of this observation for the quantization of midisuperspace models with regard to the appearence of fractional and/or emergent spacetime dimensions, as suggested by various approaches to quantum gravity. Using a simple example we show that dynamical quantum gravity effects can lead to emergent dimensions. For comparison with the results in other approaches we also calculate the spectral dimension for our model and find that the results are mutually consistent. We conclude in section V. 

\section{New second derivative theories of gravity for spherically symmetric spacetime}
In this section we present the new Hamiltonian(s) for spherically symmetric spacetimes as found in \cite{NSDTGSSS} (also see \cite{LQG-inhomogeneities, Martin-Paily}). We start by considering a spherically symmetric spacetime in $2+1$ dimensions. We will be interested in the canonical formulation of the theory and for this purpose we consider the Arnowitt-Deser-Misner (ADM) metric
\be
\md s^{2}=-N^{2}\md t^{2}+\grr(\md r+\nr\md t)^{2}+\gpp\md\phi^{2},
\ee 
where $N$ and $\nr$ are the lapse function and the shift vector respectively and $(\grr, \gpp)$ are the dynamical variables. However, for our purposes it would be more convenient to trade-off the metric coefficients $\grr$ and $\gpp$ for two new variables $\ep=\sqrt{\grr\gpp}$ and $\er=\gpp$ so that the metric takes the form  
\be \label{adm metric}
\md s^{2}=-N^{2}\md t^{2}+\frac{(\ep)^{2}}{\er}(\md r+\nr\md t)^{2}+\er\md\phi^{2}.
\ee
Because of the assumption of spherical symmetry, all these are functions of the coordinate time $t$ and the radial coordinate $r$ only (and for the same reason only the $r$-component of the shift vector $N^{i}$ is non-zero).  

The Hamiltonian and the diffeomorphism constraints corresponding to the metric \eqref{adm metric} (obtained after performing Legendre transformation on the Einstein-Hilbert Lagrangian $L=\int\md x\sqrt{-g}R$) are
\bea \label{hamiltonian constraint 2+1 dimensions}
H[N] &=& \int\md r N\bigg(-8G_{3}\kp\kr\er-4G_{3}\kp^{2}\ep \nonumber \\
&&-\frac{\epp\erp}{8G_{3}(\ep)^{2}}+\frac{\erpp}{8G_{3}\ep}-\frac{\Lambda_{3}\ep}{4G_{3}}\bigg)\approx0, \\
\label{diffeomorphism constraint 2+1 dimensions}
D[\nr] &=& \int\md r\nr(\kr\erp-\kp'\ep)\approx0.
\eea
In the above expressions $(\kp,\kr)$ are the canonical momenta conjugate to $(\ep,\er)$ respectively, and these obey the Poisson bracket relations $\{\ep(x),\kp(y)\}=\delta(x,y)$ and $\{\er(x),\kr(x)\}=\delta(x,y)$. $G_{3}$ is Newton's constant for $2+1$ dimensions while $\Lambda_{3}$ is the cosmological constant (and for reasons that will become apparent in the next section, we distinguish the $2+1$ dimensional cosmological constant by use of the subscript denoting spacetime dimensions). A prime ($'$) in the above expressions denotes a derivative with respect to $r$. 

These constraints satisfy the following Poisson bracket algebra 
\bea \label{hh constraint algebra} 
\{H[N],H[M]\} &=& D[\er(\ep)^{-2}(NM^{-1}-N^{-1}M)], \\
\label{dh constraint algebra} 
\{D[\nr],H[N]\} &=& H[N'\nr], \\
\label{dd constraint algebra}
\{D[\nr],D[M^{r}]\} &=& D[\nr M^{r'}-\nrp M^{r}].
\eea
As is well known, the above algebra encodes the diffeomorphism invariance of general relativity in the canonical formulation (specifically, diffeomorphisms in the $t-r$ plane for spherically symmetric spacetimes) \cite{Hojman-Kuchar-Teitelboim}. 

Following \cite{LQG-inhomogeneities,NSDTGSSS,Martin-Paily} we can construct many more theories (actually infinite) which are quadratic in momenta and involve at most the second derivative of the metric variables and which satisfy the \emph{same constraint algebra} as above. That is, these theories do not involve any higher derivative/curvature modifications of the Einstein-Hilbert Lagrangian and yet are not the symmetry reduced version of general relativity and at the same time satisfy the constraint algebra \eqref{hh constraint algebra}-\eqref{dd constraint algebra}. 

In brief, the basic idea for obtaining the new Hamiltonians is to first modify the Hamiltonian constraint \eqref{hamiltonian constraint 2+1 dimensions} with $\er$ dependent functions $\alpha_{i}(\er)$ so that
\bea \label{modified hamiltonian constraint 2+1 dimensions}
\bar{H}[N] &=& \int\md r N\bigg(-8G_{3}\aone\kp\kr\er-4G_{3}\at\kp^{2}\ep \nonumber \\
&&-\frac{\ath\epp\erp}{8G_{3}(\ep)^{2}}+\frac{\afo\erpp}{8G_{3}\ep}-\frac{\af\Lambda_{3}\ep}{4G_{3}}\bigg),
\eea
leaving the diffeomorphism constraint \eqref{diffeomorphism constraint 2+1 dimensions} unmodified. The next step is to evaluate the Poisson bracket
\begin{widetext}
\bea
\{\bar{H}[N],\bar{H}[M]\}&=&\int\md r\frac{\er}{(\ep)^{2}}(N'M-NM')\bigg[\aone(\afo\kp'\ep-\ath\kr\erp)+\aone(\afo-\ath)\kp\epp \nonumber \\
&&+\left(\frac{\aone\afo-\at\ath}{\er}+\afo\frac{\md\aone}{\md\er}-\aone\frac{\md\afo}{\md\er}\right)\kp\ep\erp\bigg].
\eea
\end{widetext}
For anomaly-free algebra we get the condition $\ath=\afo$ and $(\aone\afo-\at\ath)+\er(\afo\md\aone/\md\er-\aone\md\afo/\md\er)=0$. Use of the first condition already implies that the $\{D[\nr],\bar{H}[N]\}$ bracket has the standard form \eqref{dh constraint algebra}. Also the first condition when used in the second condition allows $\at$ (say) to be expressed in terms of $\aone$ and $\ath$. That is, we are left with only three arbitrary functions $(\aone,\ath,\af)$ instead of the five we started with. Use of these conditions further implies that the bracket
\be 
\{\bar{H}[N],\bar{H}[M]\}=D[\aone\ath\er(\ep)^{-2}(NM'-N'M)].
\ee

The next step is to perform a canonical transformation taking $(\er,\kr)\rightarrow(\erb,\krb)$ using the generating function $F_{3}=-\aone\ath\er\krb$. Finally, one writes the Hamiltonian $\bar{H}[N]$ and the diffeomorphism constraint \eqref{diffeomorphism constraint 2+1 dimensions} in terms of the new pair $(\krb,\erb)$. It turns out that the diffeomorphism constraint retains its form even in terms of the new variables:
\be \label{diffeomorphism constraint new variables}
D[\nr]=\int\md r\nr(\krb\erbp-\kp'\ep)\approx0.
\ee
It is also easy to check that the Hamiltonian \eqref{modified hamiltonian constraint 2+1 dimensions} written in terms of the new variables:
\begin{widetext}
\be \label{hamiltonian constraint new variables} 
\bar{H}[N]=\int\md r~N\bigg[-8G_{3}A_{1}\kp\krb\erb-4G_{3}A_{2}\kp^{2}\ep-\frac{\epp\erbp}{8G_{3}A_{1}(\ep)^{2}}+\frac{\erbpp}{8G_{3}A_{1}\ep}-\frac{A_{3}(\erbp)^{2}}{4G_{3}\ep}-\frac{A_{5}\Lambda_{3}\ep}{4G_{3}}\bigg]\approx 0, 
\ee
\end{widetext}
along with the expression \eqref{diffeomorphism constraint new variables} for the diffeomorphism constraint satisfies the constraint algebra \eqref{hh constraint algebra}-\eqref{dd constraint algebra} (see \cite{NSDTGSSS} for a detailed discussion). In the above equation $A_{1}$, $A_{2}$ and $A_{5}$ are independent and arbitrary functions of $\erb$ (which have been traded for the original and equally arbitrary functions $(\aone,\ath,\af)$) and $A_{3}$ is determined in terms of $A_{1}$ and $A_{2}$ by the relation
\be \label{A3}
A_{3}(\erb)=\frac{1}{4A_{1}\erb}+\frac{1}{2A_{1}^{2}}\frac{\md A_{1}}{\md\erb}-\frac{A_{2}}{4A_{1}^{2}\erb}.
\ee
Furthermore, it has to be remembered that now the ADM metric \eqref{adm metric} is written with $\er$ replaced by $\erb$. An explicit demonstration that the resulting theory is diffeomorphism invariant can be found in \cite{LQG-inhomogeneities} (although the fact that the constraint algebra closes in exactly the same way as for the symmetry reduced general relativity is a sufficient proof).

The presence of arbitrary functions $(A_{1}(\erb),A_{2}(\erb),A_{5}(\erb))$ implies that, in general, the theory is non-Einsteinian (it is Einsteinian in $n+2$ dimensions for specific values of these functions -- see equations \eqref{mapping a1}-\eqref{mapping a5} below; in $2+1$ dimensions this corresponds to $A_{1}=A_{2}=A_{5}=1$ and $A_{3}=0$). Actually the conclusion that the resulting models are not symmetry reduced versions of general relativity is slightly more subtle than this and following \cite{{Hojman-Kuchar-Teitelboim}} one has to verify that the momenta conjugate to the metric variables in the two theories do not differ just by terms depending on the spatial geometry but also on how the spatial slice is embedded in the spacetime if the theories given by \eqref{hamiltonian constraint new variables} are to qualify as new theories inequivalent to symmetry reduced version of general relativity (a proof of this can be found in \cite{NSDTGSSS}). 

Since for $A_{1}$, $A_{2}$ and $A_{5}$ not equal to one, the theory is not the symmetry reduced version of general relativity, one might ask what is the meaning of these models? We take the point of view that these new theories can be thought of as incorporating certain quantum gravity effects at an effective level (in the form of arbitrary functions $A_{i}$ whose exact form will be given by the underlying quantum theory of gravity). Even when not incorporating all possible quantum gravity corrections to the symmetry reduced model, these functions will nevertheless be representative of certain class of quantum gravity corrections. The presence of free functions in the Hamiltonian \eqref{hamiltonian constraint new variables} might suggest that the quantum theory is not highly constrained. However, all models of LQG analyzed using the effective-dynamics approach (as opposed to, e.g., hybrid loop quantum cosmology quantization or the dressed-metric approach to quantization), including cosmological perturbations, have such free functions \cite{QuantAmbiguity, MartinLivRev, MartinRev} (where such corrections occur quite naturally).

As an aside we would like to note that the fact that for spherically reduced $2+1$ dimensional general relativity $A_{3}=0$ means that the term involving $A_{3}$ (the $(\erbp)^{2}/\ep$ term in \eqref{hamiltonian constraint new variables}) is completely new (compare with \eqref{hamiltonian constraint 2+1 dimensions}). This is unlike what happens when one obtains new theories starting with the symmetry reduced version of $3+1$ dimensional general relativity. In that case the \emph{basic} terms in the new Hamiltonian continue to be the same as those in the Hamiltonian of the Einsteinian theory, with only the coefficients of these terms being arbitrary (similar to $A_{i}(\erb)$ here) \cite{NSDTGSSS}. 

The equations of motion $\dot{Q}=\{Q,\bar{H}[N]+\bar{D}[\nr]\}$ (where $Q \equiv (\erb,\krb,\ep,\kp)$) resulting from \eqref{diffeomorphism constraint new variables} and \eqref{hamiltonian constraint new variables} are:

\begin{widetext}
\bea \label{erdot}
\erbd &=& -8G_{3}NA_{1}\kp\erb+\nr\erbp, \\
\label{epdot}
\epd &=& -8G_{3}NA_{1}\krb\erb-8G_{3}NA_{2}\kp\ep+\nrp\ep+\nr\epp, \\
\label{kpdot}
\kpd &=& 4G_{3}NA_{2}\kp^{2}-\frac{N(\erbp)^{2}}{16G_{3}A_{1}\erb(\ep)^{2}}+\frac{NA_{2}(\erbp)^{2}}{16G_{3}A_{1}^{2}\erb(\ep)^{2}}-\frac{N'\erbp}{8G_{3}A_{1}(\ep)^{2}}+\nr\kp'+\frac{NA_{5}\Lambda_{3}}{4G_{3}}, \\
\label{krdot}
\krbd &=& 8G_{3}NA_{1}\kp\krb+8G_{3}N\kp\krb\erb\frac{\md A_{1}}{\md\erb}+4G_{3}N\kp^{2}\ep\frac{\md A_{2}}{\md\erb}-\frac{N''}{8G_{3}A_{1}\ep}+\frac{N'\epp}{8G_{3}A_{1}(\ep)^{2}} \nonumber \\
&&-\frac{N'\erbp}{8G_{3}A_{1}\ep\erb}+\frac{N(\erbp)^{2}}{16G_{3}A_{1}\ep(\erb)^{2}}-\frac{NA_{2}(\erbp)^{2}}{16G_{3}A_{1}^{2}\ep(\erb)^{2}}+\frac{N'A_{2}\erbp}{8G_{3}A_{1}^{2}\ep\erb}-\frac{N\erbpp}{8G_{3}A_{1}\ep\erb} \nonumber \\
&&+\frac{NA_{2}\erbpp}{8G_{3}A{1}^{2}\ep\erb}+\frac{N\epp\erbp}{8G_{3}A_{1}(\ep)^{2}\erb}-\frac{NA_{2}\epp\erbp}{8G_{3}A_{1}^{2}(\ep)^{2}\erb}+\frac{N(\erbp)^{2}}{16G_{3}A_{1}^{2}\ep\erb}\frac{\md A_{1}}{\md\erb}+\frac{N(\erbp)^{2}}{16G_{3}A_{1}^{2}\ep\erb}\frac{\md A_{2}}{\md\erb} \nonumber \\
&&-\frac{NA_{2}(\erbp)^{2}}{8G_{3}A_{1}^{3}\ep\erb}\frac{\md A_{1}}{\md\erb}+\frac{N\Lambda_{3}\ep}{4G_{3}}\frac{\md A_{5}}{\md\erb}+\nrp\krb+\nr\krb'.
\eea 
\end{widetext}

The presence of arbitrary functions $(A_{1},A_{2},A_{5})$ along with the new term in the Hamiltonian suggests the possibility that using this freedom one could solve the constraints \eqref{diffeomorphism constraint new variables} and \eqref{hamiltonian constraint new variables} and the equations of motion to obtain asymptotically flat static black hole solutions even for $2+1$ dimensional spacetimes - a possibility which does not exist in $2+1$ dimensional general relativity. We next show that this is a subtle issue related to the meaning of dimensionality of spacetime for symmetry reduced models.

\section{Meaning of dimensionality of spacetime for symmetry reduced models?}  

Since we started with modifications to the classical (spherically symmetric) general relativity in $2+1$ dimensions, it would appear that the modified theory is for $2+1$ dimensions. That this is not the case can be demonstrated easily. Consider the following choice for $A_{1}$, $A_{2}$ and $A_{5}$ \footnote{The coefficients $A_{i}$ in \eqref{a's for non trivial mass parameter} which lead to the classical Schwarzschild metric in $4+1$ dimensions (see \eqref{black hole solution with non trivial mass parameter}) do not correspond to the choice $n=3$ in equations \eqref{mapping a1}-\eqref{mapping a5} below. The non-trivial choice is made for two related reasons. If we choose $A_{i}$'s as in \eqref{mapping a1}-\eqref{mapping a2} with $n=3$ (and with $A_{5}=0$ so that the cosmological term in \eqref{hamiltonian constraint new variables} does not contribute since we want asymptotically flat solution), the choice would trivially correspond to general relativity though for different dimensionality of the spacetime compared to what we started with (which was $2+1$). The choice in \eqref{a's for non trivial mass parameter} demonstrates (though, admittedly, as an extreme case) the non-triviality of the notion of spacetime dimension when corrections to general relativity are present.}

\be \label{a's for non trivial mass parameter}
A_{1}(\erb)=\frac{1}{\Lambda_{3}\erb}, \quad A_{2}(\erb)=0, \quad A_{5}(\erb)=1,
\ee
which, from \eqref{A3}, implies that $A_{3}(\erb)=-\Lambda_{3}/4$. With this choice the (static) solution of the equations of motion and the constraints \eqref{diffeomorphism constraint new variables}, \eqref{hamiltonian constraint new variables} (in the gauge $\erb=r^{2}$) gives the metric 
\be \label{black hole solution with non trivial mass parameter}
\md s^{2}=-\left(1-\frac{c}{r^{2}}\right)\md t^{2}+\left(1-\frac{c}{r^{2}}\right)^{-1}\md r^{2}+r^{2}\md\phi^{2}.
\ee
We see that for $c>0$ this metric corresponds to an asymptotically flat black hole spacetime with the horizon at $r=\sqrt{c}$ (interestingly, any reference to the cosmological constant has disappeared in the metric above).

From the metric \eqref{black hole solution with non trivial mass parameter} it is obvious that this is not a solution in $2+1$ dimensions as one would have naively thought but corresponds to the Schwarzschild black hole in $4+1$ dimensions (for a fixed value of the other two angular coordinates $(\xi,\theta)$ that would be present in $4+1$ dimensions). Apart from showing that the modified theory of the previous section does not correspond to a $2+1$ dimensional theory, this example also brings into question the meaning of spacetime dimensionality for these theories.   

To answer this we recall that after imposing spherical symmetry on spacetime, we are effectively left with a theory  in the two dimensional $t-r$ plane only (as far as the spacetime dependence of the metric functions is concerned) as can be seen from the metric for spherically symmetric spacetimes: 
\be \label{adm metric for higher dimensional spherical spacetime} 
\md s^2 = -N^2 \md t^2 + \frac{(\ep)^{2}}{\er}(\md r+N^r \md t)^2 + \er \md\Omega_n^2.
\ee
The only place where the information about the number of spacetime dimensions appears is in the $\md\Omega_{n}^{2}$ part of the metric and this part is passive as far as the spherically reduced theory is concerned. 

We can, in fact, explicitly demonstrate that the information about the dimensionality of the full spacetime is lost for symmetry reduced models (this, as we will try and argue in the next section, is especially relevant in situations where the classical theory could be modified due to the presence of quantum gravity effects). We start by considering the Hamiltonian of spherically symmetric general relativity in $n+2$ dimensions (where $n$ is the number of angular dimensions and the $2$ corresponds to one radial and one temporal dimension):  
\begin{widetext}
\bea \label{classical hamiltonian n+2 dimensions}
H_{n+2} &=& \int\md r N\bigg[-\frac{\tilde{G}_{d}\kp\kr}{n(\er)^{(n-3)/2}}+\frac{(n-3)\tilde{G}_{d}\kp^{2}\ep}{4n(\er)^{(n-1)/2}}-\frac{n(n-1)\ep(\er)^{(n-3)/2}}{\tilde{G}_{d}}+\frac{n\erpp(\er)^{(n-1)/2}}{\tilde{G}_{d}\ep} \nonumber \\
&&+\frac{n(n-1)(\erp)^{2}(\er)^{(n-3)/2}}{4\tilde{G}_{d}\ep}-\frac{n\erp\epp(\er)^{(n-1)/2}}{\tilde{G}_{d}(\ep)^{2}}-\frac{2\Lambda_{d}\ep(\er)^{(n-1)/2}}{\tilde{G}_{d}}\bigg].
\eea
\end{widetext}
Here $\tilde{G}_{d}=4nG_{d}\Gamma((n+1)/2)/(n-1)\pi^{(n-1)/2}$, $G_{d}$ being Newton's constant in $d=n+2$ dimensional spacetime. There is an apparent problem with this definition for $n=1$ ($d=3$), and in that case we simply use the definition $\tilde{G}_{3}=8G_{3}$.

It turns out that the Hamiltonian in \eqref{hamiltonian constraint new variables} has enough freedom in the form of functions $A_{1}$, $A_{2}$ and $A_{5}$ that it can always be mapped to \eqref{classical hamiltonian n+2 dimensions}. If we make the choice 
\bea \label{mapping a1}
A_{1} &=& \frac{\tilde{G}_{d}}{8nG_{3}(\er)^{(n-1)/2}}, \\
\label{mapping a2}
A_{2} &=& \frac{(3-n)\tilde{G}_{d}}{16nG_{3}(\er)^{(n-1)/2}}, 
\eea
\be \label{mapping a5}
A_{5}=\frac{4n(n-1)G_{3}(\er)^{(n-3)/2}}{\tilde{G}_{d}\Lambda_{3}}+\frac{8\Lambda_{d}G_{3}(\er)^{(n-1)/2}}{\tilde{G}_{d}\Lambda_{3}},
\ee
then equation \eqref{hamiltonian constraint new variables} gets mapped to \eqref{classical hamiltonian n+2 dimensions}. Here we have distinguished $\Lambda_{3}$ from $\Lambda_{d}$ to keep things explicit. This explicitly demonstrates that from the perspective of the reduced theory one cannot say how many spacetime dimensions one is in. 

As already mentioned, this is not very surprising since for symmetry reduced models the dimensionality of the phase space is not related to the dimensionality of the embedding spacetime. For symmetry reduced classical general relativity this does not make much difference, since the information about the dimensionality of the full spacetime can be read from the coefficients in the Hamiltonian \eqref{classical hamiltonian n+2 dimensions} or from the form of the solutions of the corresponding Einstein equations which explicitly depend on the number of angular dimensions $n$. 

What is important to realize is that when one considers quantum corrections to (symmetry reduced) general relativity then, in the effective theory, the coefficients $A_{i}$ would not, in general, be the ones given by \eqref{mapping a1}-\eqref{mapping a5} but would be determined by the underlying quantum theory (here we are taking the view point that some aspects of the dynamics of quantum gravity are captured in the coefficients $A_{i}$ without the need for introducing higher derivative/curvature terms). In such a situation the \emph{classical} meaning of dimensionality of spacetime becomes non-trivial (and, in fact, need not even apply) since the corresponding solution of the equations of motion need not correspond to a solution in classical general relativity and can even lead to a notion of emergent spacetime dimensions even at the effective level as we try to demonstrate next.

\section{Possible implication(s) of the new Hamiltonians}

The choice \eqref{mapping a1}-\eqref{mapping a5} corresponds to one particular choice for $(A_{1},A_{2},A_{5})$. However, as mentioned earlier, these functions are completely arbitrary functions of $\erb$ and other choices for these functions will lead to corresponding solution metric such that, in general, the spacetime dimensionality inferred from them will be different from the dimensionality of the classical (and symmetry reduced) theory one started with (we repeat that, although presented as classical theories, the main point is that the actual form of functions $A_{i}$ will be supplied by an underlying quantum theory, see below). 

To be more specific, for $n\in\mathbb{Z}$ in equations \eqref{mapping a1}-\eqref{mapping a5}, the form of the solution will be $(1-c/r^{n-1})$. Its correspondence with the solution of the symmetry reduced general relativity then allows one to say what is the \emph{actual} dimensionality of the spacetime corresponding to which the metric is a solution (irrespective of the fact that one considered modifications to the Hamiltonian in $2+1$ dimensions), the actual dimensionality of spacetime is $d=n+2$. However, for non-integral $n$ the solution metric can have the form of the Schwarzschild solution $(1-c/r^{a})$ but with $a\in\mathbb{R}$ which can therefore be interpreted as a black hole in a spacetime with fractional dimensions! 

More generally, since we are taking the point of view that the $A_{i}$'s are given by the underlying quantum theory of gravity, the form of the functions $A_{i}$ will not be that given in \eqref{mapping a1}-\eqref{mapping a5} (even with non-integral $n$) and will additionally depend on the Planck length $\mathrm{\ell}_{P}$. Furthermore, we expect the solution incorporating quantum corrections to go over to the corresponding classical solution in the limit when the Planck length $\mathrm{\ell}_{P}\rightarrow0$ and, therefore, the semi-classical solution can be expected to have the form $(1-cf(r)/r^{n})$ (where $f(r)$ depends on the exact form of the $A_{i}$'s and will satisfy $f(r)\rightarrow1$ when $\mathrm{\ell}_{P}\rightarrow0$) instead of $(1-c/r^{a})$ which does not have the correct classical limit for a fixed $a$. 

On the other hand, a solution of the form $(1-c/r^{a})$ is still allowed by the new theories with the correct classical limit if the exponent of $r$, instead of being a constant, is a function of $r$, that is $a\equiv a(r)$ so that the dimensionality of spacetime becomes an emergent notion \footnote{Author thanks Martin Bojowald for suggesting this possibility.} (here we are working in the gauge $\erb=r^{2}$; in general, the exponent will depend on $\erb$ instead of $r$). Since the exponent has to be dimensionless, the exact dependence on $r$ (or $\erb$) will come in the combination $a(\mathrm{\ell}_{P}/r)$ (or $a(\mathrm{\ell}_{P}^{2}/\erb)$). Below we give a simple illustration of how such a scenario can be realized for a suitable choice of the functions $A_{i}$. 

Consider the following choice for the arbitrary functions (we work in the gauge $\erb=r^{2}$ and therefore write $A_{i}$'s as functions of $r$ instead of $\erb$)
\bea \label{choice of arbirary functions for emergent spacetime}
A_{1}(r) &=& (\sqrt{\Lambda}r)^{-a(r)}, \quad a(r)=m-b\frac{\ell_{P}}{r}, \quad (m\in\mathbb{N}, b\in \mathbb{R}), \nonumber \\
A_{2}(r) &=& (\sqrt{\Lambda}r)^{-a(r)}\left[1-\frac{a(r)}{2}-\frac{ra'(r)}{2}\mathrm{ln}(\sqrt{\Lambda}r)\right], \nonumber \\
A_{5}(r) &=& \frac{(\sqrt{\Lambda}r)^{a(r)-2}}{2}\left[a(r)+ra'(r)\mathrm{ln}(\sqrt{\Lambda}r)\right]. 
\eea
For this choice, the solution of equations of motion \eqref{erdot}-\eqref{krdot} and the constraints \eqref{diffeomorphism constraint new variables}, \eqref{hamiltonian constraint new variables} is
\bea 
N &=& \left(1-\frac{c\Lambda^{b\ell_{P}/2r}}{r^{a(r)}}\right)^{1/2}, \\
\ep &=& r\left(1-\frac{c\Lambda^{b\ell_{P}/2r}}{r^{a(r)}}\right)^{-1/2}.
\eea
This solution implies that the metric is 
\be \label{metric emergent spacetime}
\md s^{2}=-\left(1-\frac{c\Lambda^{b\ell_{P}/2r}}{r^{a(r)}}\right)\md t^{2}+\left(1-\frac{c\Lambda^{b\ell_{P}/2r}}{r^{a(r)}}\right)^{-1}\md r^{2}+r^{2}\md\Omega^{2}.
\ee
Since $a(r)=m-b\ell_{P}/r$ with $m\in\mathbb{N}$, we see that in the classical limit where $\ell_{P}/r\rightarrow0$, we recover the classical metric in $m+3$ spacetime dimensions (if we write $m=n-1$ then in the notation of the previous sections we have the classical metric in $n+2$ spacetime dimensions). However, away from the classical limit we find that the spacetime dimensionality will be different and, in general, will be fractional. Thus, we have here a very simple illustration of how the notion of emergent spacetime can arise from dynamical quantum gravity effects.

To make the idea of emergent dimensions arising from quantum dynamics more concrete and also to make contact with the notion of emergent dimensions used in the literature \cite{Loll-CDT, ReuterFractalAsym, ReuterLivRev, HoravaSpectralHL, CalcagniDiffusion, CalcagniSpectralLQG, CalcagniDimFlow, ModestoFractal, ModestoFractalSpinfoam, CarlipSmallStr, CarlipSpontaneous, BenedettiSpectral}, we would now like to calculate the spectral dimension for the above model. The spectral dimension $d_{\mathrm{S}}$ of a space is defined as
\be \label{spectral dimension definition}
d_{\mathrm{S}}=-2\frac{\md\,\mathrm{ln}\,P(\tau)}{\md\,\mathrm{ln}\,\tau}.
\ee
In the above equation, $P(\tau)$ is the average return probability for a diffusion process in (fictitious) `time' $\tau$, with the average return probability itself given by 
\be \label{return probability}
P(\tau)=\frac{1}{V(g)}\int\md{\bf x}\sqrt{g}\,\rho({\bf x,x};\tau)\equiv\frac{1}{V(g)}\mathrm{Tr}\,\rho(\tau),
\ee
where $\rho({\bf x, x'};\tau)$ is the probability density for diffusion from point ${\bf x}$ to point ${\bf x'}$ in time $\tau$, $g$ is the determinant of the metric on the spacetime under consideration and $V(g)$ being the corresponding volume. The probability density satisfies the heat equation 
\be \label{heat equation}
\left(\frac{\partial\,}{\partial\tau}-\square_{x}\right)\rho({\bf x,x'};\tau)=0,\quad \rho({\bf x, x'};0)=\delta({\bf x-x'}),
\ee
and is thus identified as the heat kernel ($\square_{x}$ is the Laplace-Beltrami operator on the space under consideration).

Having defined the spectral dimension we now calculate the same for the Euclideanized version of \eqref{metric emergent spacetime}. We would be interested in processes with short diffusion time $\tau$ and therefore consider the Seeley-Dewitt expansion of the heat kernel (valid for short $\tau$) according to which the trace of the heat kernel is given by (see \cite{MukhanovBook}, for instance)
\be \label{trace heat kernel}
\mathrm{Tr}\,\rho(\tau)=\int\frac{\md^{4}x\sqrt{g}}{(4\pi\tau)^{2}}\bigg(1+\tau\frac{\mathcal{R}}{6}\bigg),
\ee
where $\mathcal{R}$ is the Ricci scalar for the spacetime. Calculating the trace of the heat kernel for the Euclideanized version of \eqref{metric emergent spacetime} and using the result in eqs. \eqref{return probability} and \eqref{spectral dimension definition} one finds (we integrate in the radial direction from the origin to $R$)
\be \label{spectral dim for the model}
d_{\mathrm{S}}=4\bigg(1-\frac{5f(R)\tau}{2}\bigg),
\ee
where
\[
f(R)=\frac{c\Lambda^{b\ell_{P}/2R}}{4R^{4-b\ell_{P}/R}}(2b\ell_{P}+2R-b\ell_{P}\mathrm{ln}\Lambda R^{2}).
\]

As we do not have an underlying theory of quantum gravity, we do not expect the result to be valid for $R\approx\ell_{P}$. However, if we consider $R$ to be sufficiently large compared to the Planck length but small compared to macroscopic scales, the above expression for spectral dimension can be used to see how the spectral dimension behaves as a function of $R$. First we notice that for $R$ in the above range, $f(R)>0$ (we assume $\Lambda>0$). This implies that the spectral dimension is less than $4$. This is consistent with the results obtained for spectral dimensions at small scales in several other approaches \cite{Loll-CDT, ReuterFractalAsym, HoravaSpectralHL, CalcagniDiffusion, CalcagniSpectralLQG, CalcagniDimFlow, ModestoFractal, ModestoFractalSpinfoam, CarlipSmallStr}. Furthermore, for fixed values of $(c,\Lambda, b, \tau)$ we also find that at macroscopic scales, that is, as $R\rightarrow\infty$, $f(R)\rightarrow0$ and the spectral dimension $d_{\mathrm{S}}\rightarrow4$ and we recover the standard result.

We would like to stress that we are not saying that in quantum gravity emergent spacetime arises precisely in the manner of the previous example. That example was only for illustration purposes and the specific choice made for $a(r)$ was based on the general consideration that for $r\gg\ell_{P}$ the spacetime dimensionality should be that of the classical theory which was being quantized. The choice $a(r)=m-b\ell_{P}/r$ was one of the simplest possibilities to realize this expectation. As emphasized earlier, the exact form of $a(r)$ will be given by the underlying quantum theory of gravity. What we are pointing out is that the concept of emergent spacetime, as is suggested by various approaches to quantum gravity like the causal dynamical triangulation \cite{Loll-CDT} or fractal spacetime \cite{Calcagni-Fractal} (or even from the perspective of gauge-gravity duality \cite{Polchinski-Dualities}), can be easily accommodated even at an effective level as indicated by the example above. 

Normally quantum gravity is expected to have two implications. First, it will give rise to a microstructure for the spacetime and second, it will lead to a modified gravitational dynamics which, in general, will be important in the high curvature regime. The former is what is captured by the computation of spectral dimensions in various quantum gravity approaches while the effect of dynamical quantum gravity effects on spacetime dimensions is what we have tried to motivate in this paper. From this dynamical quantum gravity perspective and in the spirit of the example considered above, if we consider the application to black hole spacetime, then we expect that if we scatter a test particle in the background of the Schwarzschild BH (say), then in the high curvature regime the scattering behavior would be different from the classical 3+1 dimensional behavior. The particle can scatter so that it appears to be scattering in a spacetime with (effective) dimensions different from 3+1.

We would also like to add that since the main aim of the previous example was to illustrate the possibility of emergent spacetime dimensionality, we considered the simplest model of static spacetime. More realistic scenarios of emergent spacetime would most likely also have temporal evolution towards classicality and such a situation can naturally occur in the cosmological context. By incorporating suitable matter degrees of freedom in the new theories (which will lead to another $\er$-dependent arbitrary function in the matter Hamiltonian \cite{NSDTGSSS}) one can obtain dynamical models which mimic cosmological evolution but with fractional (and emergent) spacetime dimensionality at early times and go over to the classical limit only at late times (for instance, the spherically symmetric classical Lemaitre--Tolman--Bondi models, where the matter is in the form of pressureless dust, are often used to model inhomogeneous cosmology).        

It might seem that classical spacetime dimensionality (obtained from the asymptotic behavior of $g_{tt}$, say) is recovered only in the $r\rightarrow\infty$ limit, which appears unrealistic since we expect such strong quantum modifications as fractional or emergent spacetime dimensions to arise in deep quantum regime. However, as already mentioned, the specific example considered was chosen for its simplicity to bring out the key point that midisuperspace models can accommodate the scenario of emergent spacetime. In specific quantum theories, for large but finite $r$, there may still be corrections but, in general, these would keep the spacetime dimensionality close to the classical dimensions (the difference being of the order of the Planck length). Alternatively, one can view the model considered not so much as a model for black holes but rather as a model of possible space-time structures in which case the asymptotic form of the metric can be viewed as a simpler version of spectral or other dimensions used in the examples of CDT and other theories.

If one considers quantization of midisuperspace models then, demanding diffeomorphism invariance to be a good symmetry of the quantum theory one would want the quantized theory to satisfy the classical constraint algebra \eqref{hh constraint algebra}-\eqref{dd constraint algebra}. In such a situation quantum corrections would (presumably) lead to a modification of the spacetime metric (in the semi-classical regime one expects the notion of  metric to be well defined). As seen above, for a modified (effective) theory, the most natural interpretation of spacetime dimensions will probably be an emergent one, including the possibility for fractional dimensions (unless the quantum corrections \emph{only} renormalize the classical parameters like the mass and charge of the black hole, including Newton's constant, but leave the overall structure of the metric unmodified - an unlikely scenario as suggested by several studies of the effects of LQG corrections on black hole metric \cite{GambiniPullin-CompleteSpaceTime, Modesto-SpaceTimeStructure, Husain-ModifiedGR, ModifiedHorizon, Tibrewala-Einstein-Maxwell}).

As mentioned earlier, the conventional wisdom is that to have a notion of emergent spacetime, one will need to start from a non-geometrical theory like those based on matrix models or gauge-gravity duality \cite{SeibergEmergent, Polchinski-Dualities} whereas the theory we have discussed here is based on conventional geometric ideas. 
On the other hand, by now there are ample examples demonstrating the notion of emergent spacetime even in  other approaches to quantum gravity \cite{Loll-CDT, ReuterFractalAsym, ReuterLivRev, HoravaSpectralHL, CalcagniDiffusion, CalcagniSpectralLQG, CalcagniDimFlow, ModestoFractal, ModestoFractalSpinfoam, CarlipSmallStr, CarlipSpontaneous, BenedettiSpectral, AfshordiEmergentStochastic}. Additionally, we again emphasize that the viewpoint taken here is that of effective theory, in particular, the role of the modified dynamics resulting from (effective) quantum gravity. The calculations have not been performed in a full quantum gravity theory. What we have shown is that aspects of emergent spacetime might be usefully captured even at the effective level in the form of the coefficients $A_{i}$ appearing in \eqref{hamiltonian constraint new variables} whose exact form will be determined by the underlying quantum theory.

\section{Conclusions}

In this paper we highlighted that when working with symmetry reduced models of spacetime, the so called minisuperspace and midisuperspace models, one has to be careful when considering the dimensionality of spacetime in which these models are \emph{supposedly} embedded. This has to be especially so when considering the quantization of these models. Our intuition regarding the dimensionality of embeddding spacetime in based on the solutions in classical general relativity. As explicitly shown in this paper, this implies that we should allow the possibility of emergent spacetime in the quantum theory. 

Our discussion relied on the solutions obtained in recently constructed (new) second derivative theories of gravity for spherically symmetric spacetime. If we make the reasonable assumption that the classical version of the constraint algebra continues to hold even in quantum gravity (this can be thought of as an algebraic notion of general covariance) then, at least at the effective level, quantum gravity corrections will lead to a modification of the Hamiltonian of the classical theory. The new symmetry reduced models mentioned above can be seen in this light. That is, they allow possible incorporation of (dynamical) quantum gravity effects (in the form of functions $A_{i}$) without requiring the addition of higher curvature/derivative terms.

Using these models we explicitly showed that since in the presence of spherical symmetry the phase space of general relativity is of dimensions $4\infty$ irrespective of the dimensionality of the embedding spacetime, we cannot make naive conclusions about spacetime dimensions when there are quantum corrections present. To emphasize this point we also demonstrated that with a suitable choice for the functions $A_{i}$, a \emph{supposedly} $2+1$ dimensional model can be mapped to a classical(general relativistic) model in any number of spacetime dimensions. 

As a more interesting consequence of these new models we further showed that modified (quantum) gravitational dynamics allows for the possibility of emergent and fractional spacetime. To illustrate the idea we considered a particular spacetime \eqref{metric emergent spacetime} that is allowed within the class of models considered. As a concrete notion of dimensionality of (Euclideanized) spacetime we focussed on the spectral dimension $d_{\mathrm{S}}$ and found that at small scales it showed a reduction from the topological value of four. We also found that at macroscopic scales, as desired, the spectral dimension of the spacetime matched with the topological dimension. And although we considered spherically symmetric models only, we expect the conclusion to hold more generally. The possibility of emergent spacetime has been suggested by various approaches to quantum gravity \cite{Loll-CDT, CalcagniDiffusion, CalcagniSpectralLQG, CalcagniDimFlow, ModestoFractal, ModestoFractalSpinfoam, ReuterFractalAsym, ReuterLivRev, HoravaSpectralHL, Calcagni-Fractal, CarlipSmallStr, CarlipSpontaneous, SeibergEmergent, Polchinski-Dualities, BenedettiSpectral, AfshordiEmergentStochastic, Gudder-CausalSet}. For illustration purposes we considered only the case of static spacetime and the emergent nature of spacetime was apparent only with respect to the spatial scale. However, in a more realistic situation, emergent behavior in time is also expected and this can be achieved by incorporating suitable matter degrees of freedom to mimic certain cosmological scenarios.

In the context of LQG the earliest computation of spectral dimension was done in \cite{ModestoFractal} by considering the scale dependence of the area operator spectrum for spin-network states as well as for spin-foam models \cite{ModestoFractalSpinfoam}. It was found that at high energies (small scales) the spectral dimension of the spatial manifold reduces to two. A more detailed and systematic study of scale dependence of spectral dimension for quantum geometries was initiated in \cite{CalcagniSpectralLQG} where the role of the underlying discrete structures was explored by considering the full discrete Laplacian acting on (coherent) states of quantum geometry. No strong evidence for dimensional flow was, however, found. In a more recent work \cite{CalcagniDimFlow} the role of combinatorial discreteness and of superpositions of combinatorial structures was considered. It was found that, in general, the UV dimension becomes smaller than four to a state-dependent value (and for certain special kinematical states a value of $d_{\mathrm{S}}=2$ was also found).

Since some of the considerations of the present paper are motivated by LQG, it might seem worthwhile to compare the above mentioned results with those obtained here. To this end we note that although both, the more concrete LQG results of the previous paragraph and the conclusions of the presnt paper show dimensional reduction at small scales, a direct comparison of this kind would be difficult to make. The reason for this lies in certain caveats which we mention next. First of all we note that even though the construction of the Hamiltonian \eqref{hamiltonian constraint new variables} came from exploring the consequences of LQG for spherically symmetric models (see \cite{NSDTGSSS, LQG-inhomogeneities, Martin-Paily} for more details), the Hamiltonian itself is not limited to LQG corrections per se (the functions $A_{i}$ are arbitrary). Furthermore, the results of the present paper focus more on the possible role of (effective) quantum dynamics for spacetime dimensions, whereas the previous works focussed more on the kinematical aspects of the theory.

Lastly we would like to note that the particular choice for the $A_{i}$'s in \eqref{choice of arbirary functions for emergent spacetime} did not arise from LQG considerations but was made so that the resulting spacetime metric \eqref{metric emergent spacetime} has a simple (and somewhat familiar) form so that the notion of emergent spacetime dimensions is easy to appreciate. The correction functions $A_{i}$'s for LQG would be more involved but are expected to give qualitatively similar results.

\section*{Acknowledgements}
The author would like to thank Aninda Sinha for useful discussions and especially Martin Bojowald for useful discussions and for his careful reading of the manuscript and his comments on the same. The research of the author
at the Indian Institute of Science, Bangalore was supported
under the DST/1100 project of the Department of Science
and Technology, India.

\end{document}